\documentstyle[sprocl]{article}

\def\Journal#1#2#3#4{{#1} {\bf #2}, #3 (#4)}

\def\NCA{\em Nuovo Cimento}
\def\CR{\em C.R. Acad. Sci. (Paris)}
\def\CP{\em Cahiers de Physique}
\def\PRL{\em Phys. Rev. Lett.}
\def\JMP{\em J. Math. Phys.}
\def\GRG{\em Gen. Rel. Grav.}
\def\CQG{\em Class. Quantum Grav.}

\def\be{\begin{equation}}
\def\ee{\end{equation}}
\def\bea{\begin{eqnarray}}
\def\eea{\end{eqnarray}}
\def\ben{\begin{eqnarray*}}
\def\een{\end{eqnarray*}}
\font\eightmsb=msbm10 scaled 1200
\def\bbb#1{\hbox{\eightmsb#1}}

\begin{document}

\title{REMARKS ON SUPERENERGY TENSORS}

\author{JOS\'{E} M.M. SENOVILLA \footnote{Dedicated to Professor Llu\'{\i}s
Bel on the occasion of the Spanish Relativity Meeting 1998.}}

\address{Departament de F\'{\i}sica Fonamental, Universitat de Barcelona,\\
Diagonal 647, 08028 Barcelona, Spain}


\maketitle\abstracts{
We define (super)$^n$-energy tensors for non-gravitational
fields. The possibility of interchange of superenergy
between gravitational and other fields is considered.
}
  
\section{The Bel-Robinson tensor and its properties.}
The Bel-Robinson (BR) tensor\cite{B1,B2} is defined by
\begin{equation}
{\cal T}^{\alpha\beta\lambda\mu}\equiv
C^{\alpha\rho\lambda\sigma}
C^{\beta\hspace{1mm}\mu}_{\hspace{1mm}\rho\hspace{2mm}\sigma}+
\stackrel{*}{C}\hspace{.1mm}^{\alpha\rho\lambda\sigma}
\stackrel{*}{C}\hspace{.1mm}^{\beta\hspace{1mm}\mu}
_{\hspace{1mm}\rho\hspace{2mm}\sigma}   \label{BR}
\end{equation}
where $C_{\alpha\rho\lambda\sigma}$ is the Weyl tensor and * indicates the
usual dual operation: $\stackrel{*}{C}\hspace{.1mm}_{\alpha\beta\lambda\mu}
\equiv (1/2)\, \eta_{\alpha\beta\rho\sigma}
C^{\rho\sigma}_{\hspace{3mm}\lambda\mu}
=(1/2)\, \eta_{\lambda\mu\rho\sigma}C_{\alpha\beta}^{\hspace{3mm}\rho\sigma}$.
Its main properties are:

\noindent
{\bf A} It is completely symmetric and traceless
\ben
{\cal T}^{\alpha\beta\lambda\mu}={\cal T}^{(\alpha\beta\lambda\mu)},\hspace{7mm}
{\cal T}_{\hspace{2mm}\alpha}^{\alpha\hspace{1mm}\lambda\mu} =0 .
\een

\noindent
{\bf B}
For any timelike unit vector $\vec{u}$, we can define the {\it
BR super-energy {\rm (s-e)} density relative to} $\vec{u}$ as
$W_{{\cal T}}\left(\vec{u}\right)\equiv
{\cal T}_{\alpha\beta\lambda\mu}u^{\alpha}u^{\beta}u^{\lambda}u^{\mu}$.
For any $\vec{u}$ this is a positive quantity as follows from the expression
$W_{{\cal T}}\left(\vec{u}\right)    
=E_{\rho\sigma}E^{\rho\sigma}+H_{\rho\sigma}H^{\rho\sigma} \geq 0 $
where $E_{\alpha\lambda}\left(\vec{u}\right)\equiv
C_{\alpha\beta\lambda\mu}u^{\beta}u^{\mu}$ and
$H_{\alpha\lambda}\left(\vec{u}\right)\equiv
\stackrel{*}{C}_{\alpha\beta\lambda\mu}u^{\beta}u^{\mu}$
are the well-known electric and magnetic parts of the Weyl tensor, respectively
\cite{B2,Mat,B3}. Furthermore, $W_{{\cal T}}\left(\vec{u}\right)$ is zero for
some $\vec{u}$ if and only if (iff) the Weyl and BR tensors vanish
\ben
\left\{\exists \vec{u}\hspace{3mm} \mbox{such that} \hspace{2mm}
W_{{\cal T}}\left(\vec{u}\right)=0 \right\} \Longleftrightarrow
{\cal T}^{\alpha\beta\lambda\mu}=0 \Longleftrightarrow
C_{\alpha\beta\lambda\mu}=0 .
\een

\noindent
{\bf C}
The scalar $W_{{\cal T}}\left(\vec{u}\right)$ still satisfies a stronger
inequality, which has been called the {\it dominant super-energy
property} (DSEP) \cite{BS,Ber}: for all future-pointing vectors $\vec{v},
\vec{w},\vec{x},\vec{y}$, we have
${\cal T}_{\alpha\beta\lambda\mu}v^{\alpha}w^{\beta}x^{\lambda}y^{\mu}\geq 0$.
This is equivalent to the ``dominance'' of the completely timelike component
of ${\cal T}$ in any orthonormal basis $\{\vec{e}_{\mu}\}$, that is,
$W_{{\cal T}}\left(\vec{e}_0\right)={\cal T}_{0000}\geq
|{\cal T}_{\alpha\beta\lambda\mu}|$ $\forall \,
\alpha,\beta,\lambda,\mu$. This property is very useful \cite{BS} for it can
provide s-e estimates and bounds on the ``growth'' of the Weyl tensor.

\noindent
{\bf D}
Another fundamental property is that the BR tensor is ``conserved'' in
vacuum, in the sense that ${\cal T}$ is divergence-free
\ben
R_{\alpha\beta}=\Lambda g_{\alpha\beta} \hspace{1cm} \Longrightarrow \hspace{1cm}
\nabla_{\alpha}{\cal T}^{\alpha\beta\lambda\mu}=0  
\een
where $R_{\alpha\beta}$ is the Ricci tensor and $g_{\alpha\beta}$ the metric
tensor.

\section{The Bel tensor: properties and decomposition}
The BR tensor is the s-e tensor {\it par excellence}, but it was
immediately generalized by Bel to a tensor keeping most of the properties
above and valid for any gravitational field. The Bel tensor\cite{B4} is
defined by
\be
2B^{\alpha\beta\lambda\mu}\equiv \! R^{\alpha\rho\lambda\sigma}  
R^{\beta\hspace{1mm}\mu}_{\hspace{1mm}\rho\hspace{2mm}\sigma}\! +   
\!{*R*}^{\alpha\rho\lambda\sigma}
{*R*}^{\beta\hspace{1mm}\mu}_{\hspace{1mm}\rho\hspace{2mm}\sigma}\! +   
\! {*R}^{\alpha\rho\lambda\sigma}
{*R}^{\beta\hspace{1mm}\mu}_{\hspace{1mm}\rho\hspace{2mm}\sigma}\! +
\! {R*}^{\alpha\rho\lambda\sigma}
R*^{\beta\hspace{1mm}\mu}_{\hspace{1mm}\rho\hspace{2mm}\sigma}
\label{bel}
\ee
where $R_{\alpha\beta\lambda\mu}$ is the Riemann tensor and the duals act on the
left or the right pair of indices according to their
position\cite{BS2}. The properties of $B$ are:

\noindent
{\bf A} $B^{\alpha\beta\lambda\mu}=B^{\beta\alpha\lambda\mu}=    
B^{\alpha\beta\mu\lambda}=B^{\lambda\mu\alpha\beta}$,      
$B_{\hspace{1.5mm}\alpha}^{\alpha\hspace{1mm}\lambda\mu}=0$, but
$B_{\hspace{3mm}\alpha}^{\alpha\beta\hspace{1mm}\mu}\neq 0$ in general.

\noindent
{\bf B}
The Bel s-e density is defined by $W_B\left(\vec{u}\right)\equiv
B^{\alpha\beta\lambda\mu}u_{\alpha}u_{\beta}u_{\lambda}u_{\mu}$, so that
$2W_B\left(\vec{u}\right)= (EE)^2+ (HH)^2 + (HE)^2+ (EH)^2\geq 0$
where $EE$, $EH$, $HE$ and $HH$ are the electric-electric,
electric-magnetic, magnetic-electric and magnetic-magnetic parts of the
Riemann tensor, which are 2-index symmetric tensors spatial relative to
$\vec{u}$ defined by\cite{B3,BS2}
$(EE)_{\alpha\lambda}\left(\vec{u}\right)\equiv  R_{\alpha\beta\lambda\mu}
u^{\beta}u^{\mu}$, $(HH)_{\alpha\lambda}\left(\vec{u}\right)\equiv
{*R*}_{\alpha\beta\lambda\mu}u^{\beta}u^{\mu}$, 
$(HE)_{\alpha\lambda}\left(\vec{u}\right)\equiv
{*R}_{\alpha\beta\lambda\mu}u^{\beta}u^{\mu}$,
$(EH)_{\alpha\lambda}\left(\vec{u}\right)\equiv
{R*}_{\alpha\beta\lambda\mu}u^{\beta}u^{\mu}$. I use the notation
$(A)^2\equiv A_{\mu\nu\dots\rho}A^{\mu\nu\dots\rho}>0$ for any such
{\it spatial} tensor. Again, $W_B$ vanishes iff the whole Riemann and Bel
tensors vanish too
\ben
\left\{\exists \vec{u}\hspace{3mm} \mbox{such that} \hspace{2mm}
W_{B}\left(\vec{u}\right)=0 \right\} \Longleftrightarrow
B^{\alpha\beta\lambda\mu}=0 \Longleftrightarrow
R_{\alpha\beta\lambda\mu}=0 .
\een

\noindent
{\bf C}
The DSEP also holds for the Bel tensor: $B_{0000}\geq
|B_{\alpha\beta\lambda\mu}|$ for all $\alpha,\beta,\lambda,\mu$ in any
orthonormal basis. The proof of
this result is easier using spinors\cite{Ber,Bo,Ber2}.

\noindent
{\bf D}
Finally, the Bel tensor has non-vanishing divergence in general
\begin{eqnarray*}
\nabla_{\alpha}B^{\alpha\beta\lambda\mu}=
R^{\beta\hspace{1mm}\lambda}_{\hspace{1mm}\rho\hspace{2mm}\sigma}
J^{\mu\sigma\rho}+R^{\beta\hspace{1mm}\mu}_{\hspace{1mm}\rho\hspace{2mm}\sigma}
J^{\lambda\sigma\rho}-\frac{1}{2}g^{\lambda\mu}
R^{\beta}_{\hspace{1mm}\rho\sigma\gamma}J^{\sigma\gamma\rho}
\end{eqnarray*}
where $J_{\lambda\mu\beta}=-J_{\mu\lambda\beta}\equiv
\nabla_{\lambda}R_{\mu\beta}-\nabla_{\mu}R_{\lambda\beta}$. Thus, $B$ is
conserved if $J=0$.

The decomposition of the Riemann tensor into irreducible parts induces a
canonical decomposition $B^{\alpha\beta\lambda\mu}=
{\cal T}^{\alpha\beta\lambda\mu}+{\cal Q}^{\alpha\beta\lambda\mu}+
{\cal M}^{\alpha\beta\lambda\mu}$
where ${\cal Q}$ and ${\cal M}$ are called the matter-gravity and the pure
matter s-e tensors, respectively. See \cite{BS2} for their definitions
and properties. Here I only remark that ${\cal Q}$ is not typical (it does
not satisfy any of the properties A-D above) and that ${\cal M}$ is a good
s-e tensor for the matter content, it satisfies all the properties of the
Bel tensor including DSEP, and the pure-matter s-e density
$W_{{\cal M}}\left(\vec{u}\right)$ (defined as usual) is positive and vanishes
iff $R_{\mu\nu}=0$. Furthermore, the Bel s-e density decomposes as
the simple sum: $W_{B}=W_{{\cal T}}+W_{{\cal M}}$ for every $\vec{u}$.

It is noticeable that $B$ has non-zero divergence in general but is
divergence-free in the absence of matter (then $B={\cal T}$). Hence,
the question arises of how to define s-e tensors for matter fields
to see if the conservation of s-e can be restored (${\cal M}$ is not
satisfactory because it has no sense in Special Relativity).

\section{Definition of (super)$^n$-energy tensors for arbitrary physical fields}
Consider any $m$-covariant tensor $t_{\mu_1\dots\mu_m}$ as an
{\it r-fold $(n_1,\dots,n_r)$-form} (with $n_1+\dots +n_r =m$) by separating the
$m$ indices into $r$ blocks, each containing $n_A$ ($A=1,\dots,r$) completely
antisymmetric indices. This can always be done because, even if
$t_{\mu_1\dots\mu_m}$ has no antisymmetries, it can be seen as an
$m$-fold (1,\dots,1)-form. Several examples are: $F_{\mu\nu}=F_{[\mu\nu]}$ is a
simple (2)-form, while $\nabla_{\rho}F_{\mu\nu}$ is a double (1,2)-form;
the Riemann tensor is a double {\it symmetrical} (2,2)-form (the pairs can be
interchanged) and the Ricci tensor is a double symmetrical (1,1)-form;
a tensor such as $t_{\mu\nu\rho}=t_{(\mu\nu\rho)}$ is a triple symmetrical
(1,1,1)-form, etcetera. I shall denote $t_{\mu_1\dots\mu_m}$ schematically
by $t_{[n_1],\dots,[n_r]}$ where $[n_A]$ indicates the $A$-th block with $n_A$
antisymmetrical indices. Then, one can define the duals by using the * acting
on each of these blocks, obtaining the tensors (obvious notation and 4
dimensions):
\ben
t_{\stackrel{*}{[4-n_1]},\dots,[n_r]}, \dots ,\hspace{2mm}
t_{[n_1],\dots,\stackrel{*}{[4-n_r]}},\hspace{2mm}
t_{\stackrel{*}{[4-n_1]},\stackrel{*}{[4-n_2]},\dots,[n_r]}, \, \dots ,
\hspace{2mm} t_{\stackrel{*}{[4-n_1]},\dots,\stackrel{*}{[4-n_r]}} \, .
\een
There are
$2^r$ tensors in this set (including $t_{[n_1],\dots,[n_r]}$).
Contracting the first index of each block with $\vec{u}$ for all these tensors
we get the ``electric-magnetic'' decomposition of
$t_{\mu_1\dots\mu_m}$. The electric-magnetic parts will be denoted by 
\ben
(\underbrace{EE\dots E}_r)_{[n_1-1],\dots,[n_r-1]},\,\, \dots\, \dots \, ,\,
(\underbrace{E\dots E}_{r-1}H)_{[n_1],\dots,[n_{r-1}-1][3-n_r]}\, , \hspace{3mm}
\mbox{etcetera.}
\een
These $2^r$ tensors are spatial relative to $\vec{u}$ and all of them determine
$t_{\mu_1\dots\mu_m}$ uniquely and completely. Besides,
$t_{\mu_1\dots\mu_m}$ vanishes iff all its E-H parts do. Let us define the
``semi-square'' $(t_{[n_1],\dots,[n_r]}\times t_{[n_1],\dots,[n_r]})$ by
contracting all indices but one of each block in the product of $t$ with itself
\ben
(t\times t)_{\lambda_1\mu_1\dots\lambda_r\mu_r}\equiv
\prod_{A=1}^{r}\frac{1}{(n_A-1)!}\,
t_{\lambda_1\rho_2\dots\rho_{n_1},\dots ,\lambda_r\sigma_2\dots\sigma_{n_r}}
t_{\mu_1\hspace{10mm}\dots ,\mu_r}^{\hspace{2mm}\rho_2\dots\rho_{n_1}
,\hspace{5mm}\sigma_2\dots\sigma_{n_r}} \, .
\een

The {\it basic s-e tensor of $t$} is defined as the sum of the $2^r$
semi-squares constructed with $t_{[n_1],\dots,[n_r]}$ and all its duals.
Explicitly:
\bea
2T_{\lambda_1\mu_1\dots\lambda_r\mu_r}\left\{t\right\}\equiv
\left(t_{[n_1],\dots,[n_r]}\times t_{[n_1],\dots,
[n_r]}\right)_{\lambda_1\mu_1\dots\lambda_r\mu_r}+\dots \, \, \dots \, +\nonumber \\
\left(t_{\stackrel{*}{[4-n_1]},\dots,\stackrel{*}{[4-n_r]}}\times
t_{\stackrel{*}{[4-n_1]},\dots,
\stackrel{*}{[4-n_r]}}\right)_{\lambda_1\mu_1\dots\lambda_r\mu_r}
\label{set}
\eea
{\bf A}
Expression (\ref{set}) is a $2r$-covariant tensor, symmetric on each
$(\lambda_A\mu_A)$-pair, and if $t_{[n_1],\dots,[n_r]}$ is symmetric in the
interchange of $[n_A]$-blocks, then $T\{t\}$ is symmetric in the interchange
of the corresponding $(\lambda_A\mu_A)$-pairs. Moreover, $T\{t\}$ is traceless
in any $(\lambda_A\mu_A)$-pair coming from $[2]$-blocks (i.e. if $n_A=2$).
\newpage
\noindent
{\bf B} The super-energy density $W\left(\vec{u}\right)\equiv
T_{\lambda_1\mu_1\dots\lambda_r\mu_r}\{t\}
u^{\lambda_1}u^{\mu_1}\dots u^{\lambda_r}u^{\mu_r}$ is positive for
$2W\left(\vec{u}\right)=(EE\dots E)^2 +(E\dots EH)^2+\dots +(HH\dots H)^2\geq 0$,
and
\be
\left\{\exists \vec{u}\hspace{3mm} \mbox{such that} \hspace{2mm}
W\left(\vec{u}\right)=0 \right\} \Longleftrightarrow
T_{\lambda_1\mu_1\dots\lambda_r\mu_r}\{t\}=0 \Longleftrightarrow
t_{\mu_1\dots\mu_m}=0 .\label{cero}
\ee
Actually, $W\left(\vec{u}\right)$ is the sum of the squares
$|t_{\mu_1\dots\mu_m}|^2$ of all the components of $t$ in any orthonormal basis
which includes $\vec{u}=\vec{e}_0$.

\noindent
{\bf C}
The DSEP holds for the general $T\{t\}$,\cite{Ber2} which is one of the main
justifications and most important properties of definition (\ref{set}).

Example 1: \underline{The massless scalar field}. Let $\phi$ be a scalar field
satisfying $\nabla_{\mu}\nabla^{\mu}\phi =0$. First, by considering
$\nabla_{\mu}\phi$ as the basic field, one can construct the tensor
(\ref{set}) and after expanding the duals we get
\ben
T_{\lambda\mu}\{\nabla\phi\}=\nabla_{\lambda}\phi\nabla_{\mu}\phi
-\frac{1}{2}g_{\lambda\mu}\nabla_{\rho}\phi\nabla^{\rho}\phi
\een
which is the standard energy-momentum tensor. This tensor
is identically divergence-free. Second, one can use the double symmetric
(1,1)-form $\nabla_{\alpha}\nabla_{\beta}\phi$ as the basic object, and
construct the corresponding tensor (\ref{set}) which becomes (this tensor
was previously found by Bel in Special Relativity, see also \cite{BT})
\bea
T_{\alpha\beta\lambda\mu}\{\nabla\nabla\phi\}=
\nabla_{\alpha}\nabla_{\lambda}\phi \nabla_{\mu}\nabla_{\beta}\phi 
+\nabla_{\alpha}\nabla_{\mu}\phi \nabla_{\lambda}\nabla_{\beta}\phi - 
\hspace{1cm} \nonumber \\
-g_{\alpha\beta}\nabla_{\lambda}\nabla^{\rho}\phi\nabla_{\mu}\nabla_{\rho}\phi 
- g_{\lambda\mu}\nabla_{\alpha}\nabla^{\rho}\phi\nabla_{\beta}\nabla_{\rho}\phi 
+\frac{1}{2} g_{\alpha\beta}g_{\lambda\mu}
\nabla_{\sigma}\nabla_{\rho}\phi\nabla^{\sigma}\nabla^{\rho}\phi \, .
\label{sc}
\eea
This is the basic s-e tensor of the scalar field. Its divergence
can be easily computed and there appear several terms proportional to the
Riemann and Ricci tensors. Therefore, it is a {\it conserved tensor in flat
spacetime}. Now, one can go on and build the (super)$^2$-energy tensor
$T_{\alpha\beta\lambda\mu\tau\nu}\{\nabla\nabla\nabla\phi\}$ associated with
the triple (1,1,1)-form $\nabla_{\alpha}\nabla_{\beta}\nabla_{\mu}\phi$, and
so on. This produces an infinite set of basic (super)$^n$-energy tensors, one
for each natural number $n$. The following fundamental result holds: ``the
basic (super)$^n$-energy (tensor) of the scalar field vanishes iff the
(n+1)$^{th}$ covariant derivative of $\phi$ is zero''.

Example 2: \underline{The source-free electromagnetic field}. Let $F_{\mu\nu}$
be a 2-form satisfying Maxwell's equations
$\nabla_{\rho}F^{\rho}_{\hspace{-1mm}\nu}=0$,
$\nabla_{\rho}\stackrel{*}{F}{}^{\rho}_{\hspace{-1mm}\nu}=0$. The tensor
(\ref{set}) for $F$ is
\ben
T_{\lambda\mu}\{F\}=F_{\lambda\rho}F^{\rho}_{\hspace{-1mm}\mu}-
\frac{1}{4}g_{\lambda\mu}F_{\rho\sigma}F^{\rho\sigma}
\een
which is the standard divergence-free energy-momentum tensor. To define the s-e
tensor of the electromagnetic field one takes the double (1,2)-form
$\nabla_{\alpha}F_{\mu\nu}$ as basic field and construct the corresponding
expression (\ref{set})
\bea
T_{\alpha\beta\lambda\mu}\{\nabla F\}=
\nabla_{\alpha}F_{\lambda\rho}\nabla_{\beta}F_{\hspace{-1mm}\mu}^{\rho}+
\nabla_{\alpha}F_{\mu\rho}\nabla_{\beta}F_{\hspace{-1mm}\lambda}^{\rho}-
g_{\alpha\beta}\nabla_{\sigma}F_{\lambda\rho}
\nabla^{\sigma}F_{\hspace{-1mm}\mu}^{\rho}- \nonumber \\
-\frac{1}{2}g_{\lambda\mu}
\nabla_{\alpha}F_{\sigma\rho}\nabla_{\beta}F^{\sigma\rho}+
\frac{1}{4}g_{\alpha\beta}g_{\lambda\mu}
\nabla_{\tau}F_{\sigma\rho}\nabla^{\tau}F^{\sigma\rho} \, .
\label{em}
\eea
This tensor is {\it not} symmetric in the interchange of
$\alpha\beta$ with $\lambda\mu$ and is traceless in $\lambda\mu$. It does
not coincide with previous s-e tensors for $\nabla F$, such as that of
\cite{C} (which is simply $T_{\alpha\beta\lambda\mu}\{\nabla F\}+
T_{\lambda\mu\alpha\beta}\{\nabla F\}$).
Its divergence (with respect to the {\it third} index) is not zero in general,
but it vanishes {\it in flat spacetime}. Again one can construct
(super)$^n$-energy tensors associated to the higher derivatives of $F$, and the
important result is: ``the basic (super)$^n$-energy (tensor) of the
electromagnetic field vanishes iff the n$^{th}$ covariant derivative of $F$
is zero''.

Example 3: \underline{The gravitational field}. If the gravitational field
is decribed by the Riemann tensor, then the s-e tensor (\ref{set}) is
exactly the Bel tensor (\ref{bel}). Similarly, the corresponding expression
(\ref{set}) for the Weyl tensor coincides with the BR tensor (\ref{BR}). One
can also construct the s-e tensor for the Ricci tensor as basic field.
This tensor has similar properties to those of the pure-matter s-e part
${\cal M}$ of the Bel tensor, but it is {\it not} the same. In fact
\ben
2\, {\cal M}_{\alpha\beta\lambda\mu}=T_{\alpha\lambda\beta\mu}\{\tilde{R}\}+
T_{\alpha\mu\beta\lambda}\{\tilde{R}\}-T_{\alpha\beta\lambda\mu}\{\tilde{R}\}+
\frac{R^2}{72}\left(4g_{\alpha(\lambda}g_{\mu)\beta}
-g_{\alpha\beta}g_{\lambda\mu}\right)
\een
where $\tilde{R}$ indicates the trace-free Ricci tensor and $R$ is the scalar
curvature. However, it is remarkable that ${\cal M}_{\alpha\beta\lambda\mu}$
can be certainly obtained as basic s-e tensor for a field involving only
the Ricci part of the curvature, namely \cite{BS2}
\ben
{\cal M}_{\alpha\beta\lambda\mu}=T_{\alpha\beta\lambda\mu}
\left\{R_{....}-C_{....}\right\}
\een
where $R_{....}-C_{....}$ denotes the double symmetric (2,2)-form
$R_{\alpha\beta\lambda\mu}-C_{\alpha\beta\lambda\mu}$
(see e.g. \cite{BS2} for its explicit expression in terms of $R_{\mu\nu}$).
This together with the Example 2 above makes it clear that the
basic s-e tensor (\ref{set}) for a given field is {\it not} the unique one with
the same good properties.

What is the arbitrariness in the definition of s-e tensors? To answer this,
let us consider only the super-energy tensors (4-index s-e tensors). From
the definition (\ref{set}), the basic $T_{\alpha\beta\lambda\mu}$ satisfies
$T_{\alpha\beta\lambda\mu}=T_{\beta\alpha\lambda\mu}=    
T_{\alpha\beta\mu\lambda}$. The general s-e tensor formed with
index permutations exclusively (to keep DSEP) is then
\ben
\bbb{T}_{\alpha\beta\lambda\mu}\equiv c_1T_{\alpha\beta\lambda\mu}+
c_2 T_{\alpha\lambda\beta\mu}+c_3T_{\alpha\mu\lambda\beta}+
c_4T_{\lambda\beta\alpha\mu}+c_5T_{\mu\beta\lambda\alpha}+
c_6T_{\lambda\mu\alpha\beta}
\een
which does not have any symmetry in general. However, it is automatically
symmetric in the interchange of $(\alpha\beta)\leftrightarrow (\lambda\mu)$
whenever the original $T_{\alpha\beta\lambda\mu}$ has this property, in which
case we can redefine $\hat{c}_1=c_1+c_6$, $\hat{c}_2=c_2+c_5$,
$\hat{c}_3=c_3+c_4$, $\hat{c}_4=\hat{c}_5=\hat{c}_6=0$. In general, if the
original $T_{\alpha\beta\lambda\mu}$ satisfies the DSEP, then the general
$\bbb{T}_{\alpha\beta\lambda\mu}$ will also satisfy the DSEP
(and {\it a fortiori} the positivity of the super-energy and the essential
property (\ref{cero})) {\it at least} when the constants satisfy
$c_1,c_2,c_3,c_4,c_5,c_6 \geq 0$. Thus, subject to this condition there is a
six-parameter family (reduced to a three-parameter one when the interchange
between pairs holds for $T_{\alpha\beta\lambda\mu}$) of s-e tensors satisfying
the fundamental DSEP. Furthermore, $\bbb{T}_{\alpha\beta\lambda\mu}$
is symmetric in $\alpha\beta$ iff $c_2=c_4$ and $c_3=c_5$ (or
$\hat{c}_2=\hat{c}_3$ in the special case), and symmetric in $\lambda\mu$ iff
$c_2=c_3$ and $c_4=c_5$ (respectively $\hat{c}_2=\hat{c}_3$). This provides a
three-parameter (resp.\ two-parameter) family of s-e tensors with the same
symmetries as the basic one; $\bbb{T}_{\alpha\beta\lambda\mu}$ is symmetric
in the interchange of $(\alpha\beta)\leftrightarrow (\lambda\mu)$ iff $c_1=c_6$
and $c_3=c_4$ (resp., in general). In particular, this proves
that there exists a two-parameter family of s-e tensors for the gravitational
field satisfying the same properties of the Bel tensor, and another
two-parameter family for the scalar field \cite{BT}. A possible way to avoid
this arbitrariness is to take the completely symmetric part
$\bbb{T}_{(\alpha\beta\lambda\mu)}$. In the gravitational case, this provides
the s-e tensor $\bbb{T}_{(\alpha\beta\lambda\mu)}=
{\cal T}_{\alpha\beta\lambda\mu}+{\cal M}_{(\alpha\beta\lambda\mu)}$
(no matter-gravity part), which
has been also put forward recently by Robinson\cite{R}.

\section{Interchange of super-energy between different physical fields}
The important question of whether or not super-energy has any physical
reality or interest may find an answer by studying its interchange between
fields (as energy does) and its possible conservation. One of the most striking
features of the Bel tensor is that it is conserved in the absence of matter
(when it coincides with the BR tensor). Similary, the s-e tensors (\ref{sc})
and (\ref{em}) are conserved in the absence of gravity (in flat spacetime).
The natural question arises: can these tensors be combined to produce a
conserved quantity?

To answer it, let us consider the propagation of
discontinuities of the Riemann and other fields. The notation and conventions
we use are those in Ref.\cite{MS}: $\Sigma$ denotes a null hypersurface whose
first fundamental form is $\bar{g}_{ab}$, ($a,b,\dots =1,2,3$ are indices in
$\Sigma$). The normal one-form to $\Sigma$ is $n_{\mu}$ ($n_{\mu}n^{\mu}=0$),
and a basis of tangent vectors to $\Sigma$ is denoted by $\vec{e}_a$
($n_{\mu}e^{\mu}_a =0$). Obviously $\vec{n}=n^a\vec{e}_a$ and
$\bar{g}_{ab}n^a=0$. As the null vector $\vec{n}$ is in fact tangent to $\Sigma$
one needs to choose a vector field {\it transversal} to $\Sigma$ which is
called the rigging \cite{MS,Sc} and denoted by $\vec{\ell}$
($n_{\mu}\ell^{\mu}=1$). There are many different choices for $\vec{\ell}$,
but given any of them we can define $\omega^a_{\mu}$ as the three one-form
fields completing with $n_{\mu}$ the basis dual to $\{\vec{\ell},\vec{e}_a\}$,
that is, $\ell^{\mu}\omega^a_{\mu}=0$, $e^{\mu}_b\omega^a_{\mu}=\delta^a_b$.
We also put $\bar{g}^{ab}\equiv g^{\mu\nu}\omega^a_{\mu}\omega^b_{\nu}$.
Notice that this is {\it not} the inverse of $\bar{g}_{ab}$, which does not
exist because $\Sigma$ is null and the first fundamental form is degenerate
\cite{MS}. This last fact also implies that there is no canonical metric
connection associated to $\bar{g}_{ab}$ in $\Sigma$. However, given any rigging
one can define the so-called rigged connection\cite{MS,Sc} in $\Sigma$ by means
of $\bar{\Gamma}^a_{bc}\equiv \omega^a_{\rho}e^{\mu}_b \nabla_{\mu} e^{\rho}_c$,
which is obviously torsion-free ($\bar{\Gamma}^a_{bc}=\bar{\Gamma}^a_{cb}$).
The covariant derivative associated to the rigged connection will be written
as $\overline{\nabla}$. The discontinuity of any object $v$ across $\Sigma$ will
be denoted by $\left[v\right]$, as is customary. A very important point is that
all the formulae appearing in what follows are {\it independent} of the
choice of the rigging vector $\vec{\ell}$.

\underline{The vacuum case.} As is well-known, the discontinuity across $\Sigma$
of the Riemann tensor is determined by a symmetric tensor $B_{ab}$ on $\Sigma$
(in principle, six independent components), as follows
\be
\left[R_{\alpha\beta\lambda\mu}\right]=B_{ab}
\left(n_{\alpha}\omega^a_{\beta}-n_{\beta}\omega^a_{\alpha}\right)
\left(n_{\lambda}\omega^b_{\mu}-n_{\mu}\omega^b_{\lambda}\right)\, .
\label{disc}
\ee
However, in vacuum only two independent discontinuities survive because
$\bar{g}^{ab}B_{ab}=0$ and $n^aB_{ab}=0$. Then, one can show\cite{L,S}
\ben
\overline{\nabla}_a\left(B^2n^an^bn^cn^d\right)=0, 
\een
where $B^2\equiv B^{ab}B_{ab}=\bar{g}^{ac}\bar{g}^{bd}B_{ab}B_{cd}>0$. Moreover,
the object inside brackets is directly related to the BR tensor by
\ben
2B^2n^an^bn^cn^d=\omega^a_{\alpha}\omega^b_{\beta}\omega^c_{\lambda}
\omega^d_{\mu}\left[{\cal T}^{\alpha\beta\lambda\mu}\right] \, .
\een
Therefore, {\it the discontinuity of the BR tensor is conserved along $\Sigma$
in vacuum}.

\underline{The Einstein-Maxwell case.} If there is only electromagnetic field
$F_{\mu\nu}$ in the spacetime, and this is continuous, the discontinuities of
the Riemann tensor take the same form (\ref{disc}) and besides
\ben
\left[\nabla_{\lambda}F_{\mu\nu}\right]=n_{\lambda}f_{\mu\nu}, \hspace{5mm}
f_{\mu\nu}=f_a\left(n_{\mu}\omega^a_{\nu}-n_{\nu}\omega^a_{\mu}\right),
\hspace{5mm} n^af_a=0
\een
so that there only appear two independent discontinuities in the first
derivatives of $F$. It can be proved\cite{L,S} that
\be
\overline{\nabla}_a\left\{\left(B^2+f^2\right) n^an^bn^cn^d\right\}=0, 
\label{fund}
\ee
where $f^2\equiv f^af_a =\bar{g}^{ab}f_af_b>0$. The interesting thing is that
this quantity can be related to the s-e tensor (\ref{em}), {\it if} $F_{\mu\nu}$
propagates in vacuum, by
\ben
2f^2n^an^bn^cn^d=\omega^a_{\alpha}\omega^b_{\beta}\omega^c_{\lambda}
\omega^d_{\mu}\left[T^{\alpha\beta\lambda\mu}\{\nabla F\}\right] \, .
\een
Notice that in this case we have $\left[B_{\alpha\beta\lambda\mu}\right]=
\left[{\cal T}_{\alpha\beta\lambda\mu}\right]$, because the pure-matter
part is continuous. Thus, {\it if a Maxwell field propagates in vacuum, the sum
of the discontinuities of the Bel tensor and of the s-e tensor for the
electromagnetic field defined in (\ref{em}) is conserved along $\Sigma$}.
A version of this fundamental result and of relation (\ref{fund}) were
obtained by Lichnerowicz\cite{L} back in 1960.

\underline{The case with a massless scalar field.} Assume finally that there
only exists a massless scalar field $\phi$ in the spacetime whose
gradient is continuous so that the energy-momentum tensor and the pure-matter
part of the Bel tensor are also continuous,
$\left[{\cal M}_{\alpha\beta\lambda\mu}\right]=0$. Then, expression (\ref{disc})
still holds and furthermore
\ben
\left[\nabla_{\mu}\nabla_{\nu}\phi\right]=V\,n_{\mu}n_{\nu}
\een
which provides a unique discontinuity freedom. Then, one can arrive at\cite{S}
\ben
\overline{\nabla}_a\left\{\left(k_1B^2+k_2V^2\right) n^an^bn^cn^d\right\}=0,
\hspace{1cm} \forall k_1,k_2
\een
and where, analogously as before, if the scalar field vanishes at one side of
$\Sigma$
\ben
2V^2n^an^bn^cn^d=\omega^a_{\alpha}\omega^b_{\beta}\omega^c_{\lambda}
\omega^d_{\mu}\left[T^{\alpha\beta\lambda\mu}\{\nabla \nabla \phi\}\right] 
\een
for the s-e tensor defined in (\ref{sc}). However, in this case, given
that $k_1$ and $k_2$ are arbitrary constants, the discontinuities are
divergence-free separately. In other words, {\it under the stated assumptions,
the discontinuities of the Bel tensor and of the s-e
tensor for the scalar field defined in (\ref{sc}) are conserved along $\Sigma$
independently}.

Similar studies can be carried out for the higher-order (super)$^n$-energy
tensors. The relevance and interpretation of all these results are under
current investigation.

\section*{Acknowledgements}
I am grateful to Llu\'{\i}s Bel, G\"oran Bergqvist, and Pierre Teyssandier for
helpful discussions, useful comments and important information.

\end{document}